\documentclass{aa}  

\usepackage{graphicx}
\usepackage{txfonts}
\usepackage{lipsum}
\usepackage{subcaption}         
\usepackage{lscape}             
\usepackage{placeins}           
            
\usepackage[colorlinks=true, linkcolor=blue, urlcolor=blue, citecolor=blue]{hyperref}

\usepackage{scalerel}
\usepackage{tikz}
\usetikzlibrary{svg.path}
\definecolor{orcidlogocol}{HTML}{A6CE39}
\tikzset{
  orcidlogo/.pic={
    \fill[orcidlogocol] svg{M256,128c0,70.7-57.3,128-128,128C57.3,256,0,198.7,0,128C0,57.3,57.3,0,128,0C198.7,0,256,57.3,256,128z};
    \fill[white] svg{M86.3,186.2H70.9V79.1h15.4v48.4V186.2z}
                 svg{M108.9,79.1h41.6c39.6,0,57,28.3,57,53.6c0,27.5-21.5,53.6-56.8,53.6h-41.8V79.1z M124.3,172.4h24.5c34.9,0,42.9-26.5,42.9-39.7c0-21.5-13.7-39.7-43.7-39.7h-23.7V172.4z}
                 svg{M88.7,56.8c0,5.5-4.5,10.1-10.1,10.1c-5.6,0-10.1-4.6-10.1-10.1c0-5.6,4.5-10.1,10.1-10.1C84.2,46.7,88.7,51.3,88.7,56.8z};
  }
}

\newcommand\orcidicon[1]{\href{https://orcid.org/#1}{\mbox{\scalerel*{
\begin{tikzpicture}[yscale=-1,transform shape]
\pic{orcidlogo};
\end{tikzpicture}
}{|}}}}
\usepackage{hyperref}
\usepackage{booktabs}
\usepackage{amsmath,amssymb,amsfonts}
\usepackage{algorithmic}
\usepackage{textcomp}
\usepackage{mwe}
\usepackage[switch]{lineno}
\renewcommand\makeLineNumber{}

\begin{document}

   \title{Not so Swift: 20 years of multiwavelength observations of \\ Mrk\,421 and Mrk\,501}

   \author{Gabrielle L. Taylor\inst{1}\fnmsep\thanks{Fellow of the International Max Planck Research School for Astronomy and Cosmic Physics at the University of Heidelberg (IMPRS-HD).}
   \and Stefan J. Wagner \inst{1}
   \and Alicja Wierzcholska\inst{2}
   \and Michael Zacharias\inst{1,3}
   }

   \institute{Landessternwarte, Universität Heidelberg, Königstuhl, D-69117 Heidelberg, Germany \\ \email{gtaylor@lsw.uni-heidelberg.de}
   \and Institute of Nuclear Physics, Polish Academy of Sciences, PL-31342 Krakow, Poland
   \and Centre for Space Research, North-West University, Potchefstroom 2520, South Africa
   }

   \date{Received 10 October 2025; accepted 7 January 2026}

   \abstract
   {}
   {The blazars Mrk\,421 and Mrk\,501 have shown multiwavelength variability on all observed timescales, and have been well studied at high energies on short timescales. We aim to characterise the long-term temporal behaviour of these blazars at synchrotron energies, namely optical, UV, and X-ray, in order to assess current models of these objects and their processes.}
   {Amongst the longest light curves ever studied for these sources, we investigated 20 years of data (2005-2025) from the Swift-UVOT and Swift-XRT telescopes. We examined spectral models, fractional variabilities, flux distributions, and X-ray photon index vs flux relations, as well as carrying out in-depth time series analysis using structure functions, Lomb-Scargle periodograms, and discrete correlation functions.}
   {Mrk\,421 and Mrk\,501 both showed intriguing variability in all studied wavelengths; this variability has been found to be energy dependent, as has the trend of lognormality in flux distributions.
   X-ray photon indices fluctuated greatly throughout the entire period, showing an overall harder-when-brighter trend. Hints of a quasi-periodicity have been found in the X-ray of Mrk\,501 (host frame time scale $\sim$390 days, >3$\sigma$) but not in the UV or X-ray of Mrk\,421, or in the UV of Mrk\,501. No correlation at any time lag was found between the optical/UV and X-ray bands in either source.
   }
   {}


   \maketitle

\section{Introduction}

Blazars are a subclass of active galactic nuclei (AGN) with their nonthermal jets directed towards the Earth's line of sight \citep{Blandford1974}, and are characterised by high, frequency dependent variability across the entire electromagnetic spectrum. They can be further classified according to the presence of broad emission lines, where those of flat spectrum radio quasars (FSRQs) have an equivalent width $EW>5\,$\AA, and those of BL Lac objects have $EW<5\,$\AA \, \citep{Urry1995}. The broadband spectral energy distributions (SEDs) of blazars contain two peaks, the lower energy peak stretches from the radio to the UV/X-ray regime, and the higher from X-ray to the $\gamma$-ray regime. The lower energy peak is generally attributed to synchrotron radiation from relativistic electrons gyrating in the jet's magnetic field, and is therefore also called the synchrotron peak. Depending on the location of the synchrotron peak frequency ($\nu_{sp}$), BL Lac objects are divided once again into low-frequency peaked BL Lac objects (LBLs) ($\nu_{sp}<10^{14}$Hz), intermediate peaked BL Lac objects (IBLs) ($10^{14}$Hz $<\nu_{sp}<10^{15}$Hz), and high-frequency peaked BL Lac objects (HBLs) ($10^{15}$Hz$<\nu_{sp}$) \citep{Padovani1995}. 

Various emission models have been proposed to try to explain the observed blazar SEDs, all of which generally agree that the lower energy peak is produced by the aforementioned synchrotron radiation. To explain the higher frequency peak, \cite{Blandford1974} suggested the one-zone synchrotron self-Compton (SSC) model, in which the same electrons which produce the synchrotron photons of the lower frequency peak then inverse Compton scatter those same photons and upscatter them to higher energies. \cite{Dermer1992} introduced the external inverse Compton (EIC) model, in which the electrons upscatter external photons from the accretion disc. Other important external photon sources stem from the broad-line region \citep{Sikora1994} and/or the dusty torus \citep{Blazejowski2000}. These SSC and EIC scenarios are strictly leptonic, but a lepto-hadronic concept has also been suggested \citep{Mannheim1993}. In this framework, relativistic protons emit via proton-synchrotron radiation, and/or collide to create (neutral, positive, and negative) pions which rapidly decay into photons, neutrinos, electrons, and positrons. A comprehensive review of these mechanisms can be found in \cite{Bottcher2007, Cerruti2020}, and references therein. 

Blazars show puzzling variability across the entire electromagnetic spectrum \citep[e.g.,][]{Cui2004}, sometimes showing correlation between bands, but then the correlation disappears, \citep[e.g.,][]{HESS_Abramowski2014}. The one-zone SSC model implies all emission be correlated, hence cannot explain any lack of correlations. The one-zone EIC model implies the low and high frequency humps should be correlated within themselves but not with each other, so also cannot explain uncorrelated patterns. To overcome this issue, the multi-zone SSC model was introduced \citep[e.g.,][]{Graff2008}, which can effectively model the correlations, or lack thereof, in radio, optical, UV, X-ray, and $\gamma$-ray energies. Lepto-hadronic models may also be able to explain the complex correlations between the X-rays and the optical/UV and/or $\gamma$-rays.

In order to assess the validity of the one-zone vs multi-zone emission models, it is therefore paramount to study and analyse the emissions observed from blazars in the lower energy peak.
Mrk\,421 (RA$_\text{J2000}$: 11h 04m 27.314s, DEC$_\text{J2000}$: +38° 12' 31.80", $z\approx0.030$), where $z$ is redshift, and Mrk\,501 (RA$_\text{J2000}$: 16h 53m 52.21s, DEC$_\text{J2000}$: +39° 45' 37.6",  $z\approx0.034$) are two of the closest HBLs to Earth, making them excellent candidates to examine the behaviour and mechanisms of these objects. 

Both sources are well studied in the gamma regime, see for example \cite{Punch1992} for Mrk\,421 and \cite{Quinn1996} for Mrk\,501, but less so at lower energies. Some papers focus the X-ray regime but not optical/UV, e.g. \cite{Blazejowski2005, Li2016}, and \cite{Alfaro2025} for Mrk\,421 and \cite{Sambruna2000, Roedig2009, Devanand2022, Alfaro2025}, and \cite{Panis2025} for Mrk\,501. Others examine optical/UV but not X-ray, e.g. \cite{Sandrinelli2017} for Mrk\,421 and \cite{Bhatta2021} for Mrk\,501. Works which investigate both optical/UV and X-ray regimes, for example \cite{Chatterjee2021} for Mrk\,421 and \cite{Catanese1997} for Mrk\,501, are often focused on short timescales, that is, less than roughly a week.

For Mrk\,421 multiwavelength (MWL) studies including optical/UV and X-ray data over timescales ranging from 2 months to 8 years have been done by e.g. \cite{Tramacere2009, Ahnen2016, Carnerero2017, Kapanadze2016, Kapanadze2017_421, Kapanadze2018a, Kapanadze2018b, Kapanadze2020, Kapanadze2023, Kapanadze2025}, and \cite {ArbetEngels2021}. These generally agree that the optical/UV and X-rays are not correlated, hence a multi-zone model is required. None find indications of quasi-periodicities in any wavelength. \cite{Sinha2016} explored 6 years of MWL data from Mrk\,421 and found that variability is energy dependent, with strong correlations between optical and GeV bands, and between X-ray and very high energy bands. This is supportive of a one-zone SSC model. Of the aforementioned longer timescale MWL papers, those which investigate the following find that: curved models fit X-ray spectra best, variability and distributions' lognormality depend on wavelength, and X-ray spectra are harder when flux is higher.

For Mrk\,501, MWL analyses which involve optical/UV and X-ray data over timescales ranging from 8 months to 6 years have been done by, for example, \cite{Kapanadze2017_501, Kapanadze2023}, and \cite{Tantry2024}. Similarly to Mrk\,421, these studies found that X-ray spectra are best fit with curved models, variability is  energy dependent, and X-ray spectra harden with increase in flux. The former found that the lognormality of flux distributions depends on wavelength, but no evidence of periodic patterns, and no correlations between optical/UV and X-rays.

In this paper, we present and analyse 20 years of quasi-simultaneous optical/UV and X-ray data for Mrk\,421 and Mrk\,501. This large increase in the length of our light curves allows us to probe previously inaccessible timescales, and draw conclusions with greater confidence. The data cover the period from April 2005 to April 2025 ($\sim$ MJD 53430 - 60793) of Mrk\,421 and Mrk\,501 taken with the Neil Gehrels Swift Observatory (hereafter Swift) employing its UVOT \citep{Roming2005} and XRT \citep{Burrows2005} telescopes. We comment on the implications of these results in the context of the existing emission models. The analysis of the data is described in Sec.~\ref{sec:analysis}. The MWL light curves are presented in Sec.~\ref{sec:Results}, alongside their fractional variabilities, flux distributions, and photon indices. Temporal characteristics including quasi-periodic oscillations (QPOs), and models which could explain them, are also discussed in detail. Sec.~\ref{sec:concs} provides a summary of our key findings and conclusions.

\section{Data analysis} \label{sec:analysis}

\subsection{Optical/UV data}

The optical and UV observations were collected from April 2005 - April 2025 in six filters using the Swift-UVOT instrument: V (544 nm), B (439 nm), U (345 nm), UVW1 (251 nm), UVM2 (217 nm), and UVW2 (188 nm). For each observation, photon counts were extracted from a circular region with a radius of 5 arcseconds centred on the source. These are not binned light curves, instead each data point represents one UVOT observation, with an minimum exposure time of 0.5\,ks.
The background was estimated from a nearby circular region, free of contamination from other sources. Instrumental magnitudes and corresponding fluxes were derived using the \verb|uvotsource| task. Flux calibration was performed using the conversion factors provided by \citet{Poole08}. The resulting fluxes were corrected for Galactic dust extinction, adopting a colour excess of $E(B-V) =  0.0132 $mag and $E(B-V) = 0.0163$mag, for Mrk\,421 and Mrk\,501, respectively \citep{Schlafly11}, and using extinction coefficients $A_{\lambda}/E(B-V)$ from \citet{Giommi06}.

For the purpose of maximising the number of data in each light curve, the narrow-band UV data were combined into one `UV' light curve, $f_{UV}$, by finding the weighted mean of all data points within 0.1 day bins, across the three narrow-band UV filters. This is justified by the fact that the narrow filter UV bands overlap significantly and fluxes derived in the three bands are not independent, see Fig.~4 in \cite{Roming2005}. Therefore, the UV flux is calculated as follows:

\begin{equation}
    f_{UV} = \frac{(f_{W1}\times\bar{f}_{W1}) + 
                   (f_{M2}\times\bar{f}_{M2}) + 
                   (f_{W2}\times\bar{f}_{W2})}
    {\bar{f}_{W1} + \bar{f}_{M2} + \bar{f}_{W2}} 
    \label{eq:UVflux}
\end{equation}

where $f_{M1}$, $f_{M1}$, and $f_{M1}$ are the UVW1, UVM2, and UVW2 fluxes, respectively, within a given time bin. $\bar{f}_{W1}, \bar{f}_{M2}$, and $\bar{f}_{W2}$ are the mean values of the UVW1, UVM2, and UVW2 light curves. The associated error, $\sigma_{UV}$, is: 

\begin{equation}
    \sigma_{UV} = \sqrt { \frac{
            (\sigma_{W2}^2\times\bar{f}_{W1}) + 
            (\sigma_{M2}^2\times\bar{f}_{M2}) + 
            (\sigma_{W2}^2\times\bar{f}_{W2})}
    {3\times (\bar{f}_{W1} + \bar{f}_{M2} + \bar{f}_{W2})} }
    \label{eq:UVerr}
\end{equation}

where $\sigma_{W2}, \sigma_{M2}$, and $\sigma_{W2}$ are the errors of the UVW1, UVM2, and UVW2 fluxes, respectively, within a given time bin. If data were only taken in two or one of the bands in any given time bin, the numerators and denominators of Eqs.~\ref{eq:UVflux} and \ref{eq:UVerr} were modified to include only the terms corresponding with the available bands, and the constant 3 was altered to 2 or 1, respectively.

\subsection{X-Ray data}

Simultaneous X-ray observations were conducted, corresponding to the ObsIDs of 00052100001-00034228234 and 00030793149-00015411233 for Mrk\,421 and Mrk\,501, respectively. The X-ray data analysis was performed using the HEASOFT software (version 6.35), and data were recalibrated using the standard \verb|xrtpipeline| procedure. For the spectral fitting of each observations \verb|xspec| was used \citep{Arnaud}, again these are not binned light curves and each data point represents one observation with a minimum exposure time of 0.5\,ks.
In the case of each observation's spectrum, the data are binned to have at least 30 counts per energy bin. Energy fluxes have been derived in two ways: by fitting each single observation with a single power law model, with Galactic absorption values of {\it N}$_H$= 1.34$\times$10$^{20}$\,cm$^{-2}$ and {\it N}$_H$= 1.69$\times$10$^{20}$\,cm$^{-2}$ for Mrk\,421 and Mrk\,501, respectively \citep[][]{HI4PI}; and by fitting each single observation with a log parabolic model with the same values for Galactic absorption.
The Swift/XRT data have been corrected for pile-up effects if needed.

The energy range of the Swift-XRT instrument is 0.3\,keV - 10\,keV. For each X-ray observation, two models are used to describe the spectra: 

\begin{itemize}
 \item a single power law:
 \begin{equation}
\frac{dN}{dE}=N_p  \left( \frac{E}{E_0}\right)^{-{\Gamma}}
\label{eq:powerlaw}
\end{equation}
 with the spectral index, $\Gamma$, the normalisation, $N_p$, and the normalisation energy $E_0=1$\,keV; and

\item a log parabola (LP):
 \begin{equation}
\frac{dN}{dE}=N_l  \left( \frac{E}{E_0}\right)^{-({\alpha+\beta \log (E/E_0)})}
\label{eq:logparabola}
\end{equation}
 with the normalisation, $N_l$, the spectral index, $\alpha$, and the curvature parameter, $\beta$.
\end{itemize}

In order to determine the most appropriate spectral model for the X-ray emission of Mrk\,421 and Mrk\,501, for each observation we evaluated the goodness of fit and compared the two nested models using the F test \citep[e.g.][]{Bevington}. For both sources, the F test returned p-values $<10^{-5}$, indicating that the improvement in fit is highly significant with the curved LP model used, also agreeing with recent work by \cite{Alfaro2025}. This suggests that the X-ray spectra of Mrk\,421 and Mrk\,501 are intrinsically curved, and are better described by a log parabolic shape rather than a single power law.

\section{Results and discussion}
\label{sec:Results}

\subsection{Light curves and simultaneous correlations} \label{sec:LCs}

\begin{figure*}[htbp]
   \centering
   \captionsetup{width=\linewidth}
   \includegraphics[width=\linewidth]{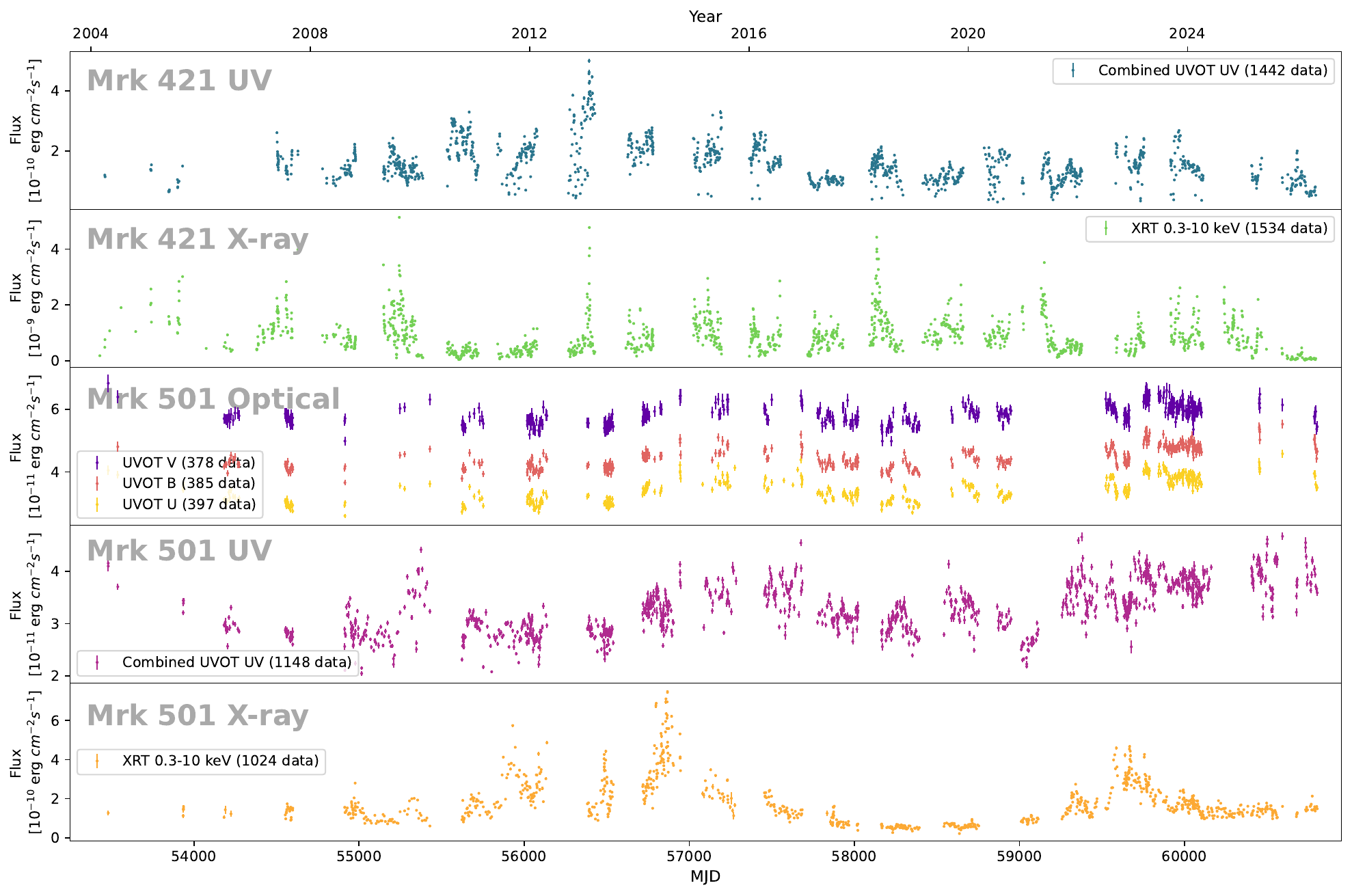}
   \caption{Multiwavelength light curves of Mrk\,421 UV, Mrk\,421 X-ray, Mrk\,501 optical, Mrk\,501 UV, Mrk\,501 X-ray. The legend shows the Swift telescope and filter used to collect the data, as well as the number of observations in each band. Each data point represents one observation.
   We present the full dynamic range of the data, in linear scaling, and observe higher variability at higher frequencies.}
   \label{fig:0_LCs}
\end{figure*}

We present the 20 year Swift-UVOT and Swift-XRT light curves of Mrk\,421 and Mrk\,501 in Fig.~\ref{fig:0_LCs}, which are amongst the longest for the sources at the time of writing. Observations were taken for Mrk\,421 between November and June each year, and for Mrk\,501 between March and September, except for the last few years where observations of Mrk\,501 have been continuous. Long and short-term variability is observed across all presented wavelengths in both sources, though more pronouncedly in the X-rays. The legend of Fig.~\ref{fig:0_LCs} shows the number of observations in each band. For Mrk\,421 there are so little data in the optical (<50 in each filter) that we do not present these light curves. Mrk\,421 appears to have reached a new and very stable low state in the X-ray band in 2025.

In both sources, strong correlations within the optical bands, and strong correlations between the optical and UV bands, are observed. Strong correlation is said to be $r\geq0.8$, with $r$ being the Pearson correlation coefficient. This is illustrated in Fig.~\ref{fig:3_Corrs}, and is expected because of the broad emissivity of photons in this wavelength range. Oppositely, no simultaneous correlations are present between the optical/UV bands and X-ray band, though this is only strictly true when considering the entire period as one. 
 However, if the X-ray data are divided by their spectral index, $\alpha$, this is no longer true; see Fig.~\ref{fig:3_Corrs}. In theory we divide at $\alpha=2$, and spectra with $\alpha<2$ are called hard whereas spectra with $\alpha>2$ are called soft. In practice we divide the populations into $\alpha>1.1\overline{\alpha}$ and $\alpha<0.9\overline{\alpha}$ for each source (where $\overline{\alpha}$ is the mean spectral index), in order to create a safe boundary between them, accounting for errors.
For Mrk\,421 the upper and lower boundary values are 2.4 and 2.0, respectively, and for Mrk\,501 they are 2.2 and 1.8, respectively. In both cases, hard X-rays vs UV and soft X-rays vs UV, we see an increase in correlation for each source, albeit on a low level. Interestingly, the magnitude of the increase in correlation is different between the sources. Specifically, in Mrk\,421 the UV are more correlated with the hard X-rays than with the soft X-rays, whereas in Mrk\,501 the UV are more correlated with the soft X-rays than with the hard X-rays. In general we may expect to see increased correlation between bands during flaring states, that is, hard X-ray spectral states, which is mildly observed in Mrk\,421 but not in Mrk\,501 \citep[for the latter see, e.g., Fig.~2 in][]{Magic2024Abe_501}, hinting at underlying differences in flare evolutions between the two sources. Notably, this is not visible when considering the whole data set. Because the correlations are far below the $r\geq0.8$ level, and because dividing the X-rays greatly reduces the number of data points in each case (because of the `boundary' between subpopulations), we do not make the index-based distinction again. Irrespective of the correlation value, the UV flux always covers the same range no matter the X-ray flux value or hardness state.

It is important to state that in these correlation plots, Fig.~\ref{fig:3_Corrs}, simultaneous observations are those taken within 0.1 days of each other. An in-depth discussion on the implications of correlations at varying time lags is in Sec.~\ref{sec:CrossCorrs}.

\subsection{Fractional variability} \label{sec:frac_var}

The fractional variability, $F_{var}$ \citep{Vaughan2003,Schleicher2019}, of each band for both blazars is given in Tab.~\ref{tab:fracvars}, and is calculated using 

\begin{equation}
    F_{var} = \sqrt{\frac{S^2 - \overline{\sigma^2}_{err}}{\bar{F}}}
\end{equation}

\noindent where $S^2$ is the variance, $\overline{\sigma^2_{err}}$ is the mean squared error, and $\bar{F}$ is the mean flux. Its error is given by

\begin{equation}
    F_{var,err} = \sqrt{\frac{1}{2N}
                        \left(\frac{\overline{\sigma^2}_{err}}{F_{var} \bar{F}}\right)^2
                        +
                        \frac{1}{N}
                        \frac{\overline{\sigma^2}_{err}}{\bar{F}}}
\end{equation}

\noindent where $N$ is the number of data points in the light curve. In both sources the fractional variability shows a clear dependence on energy (within the first SED peak), as is commonly seen in blazar observations \citep[e.g.,][]{Sinha2016, Tantry2024, Abe2025}. The X-rays have higher variability in contrast to the optical/UV photons. This statement is observed as presented in Tab.~\ref{tab:fracvars}, and remains true if we correct for the worst case contribution of the host galaxy (not presented). This could be because X-rays are produced by more energetic electrons, which have the shorter cooling time, whereas optical/UV photons come from the lower energy electron population, which takes longer to cool, and are hence less variable \citep{Tantry2024, Abe2025}.

\begin{table}[htbp]
\centering
\captionsetup{width=.95\linewidth}
\caption{Fractional variability in each filter of Mrk\,421 and Mrk\,501.}
\begin{tabular}{c c c}
 \toprule
 Filter & Mrk\,421 & Mrk\,501 \\
 \midrule
 V     & 0.257 & 0.044 \\
 B     & -     & 0.075 \\
 U     & 0.335 & 0.109 \\ 
 UV    & 0.403 & 0.143 \\
 X-ray & 0.737 & 0.615 \\
\bottomrule
\end{tabular}
\tablefoot{The errors are of the order 0.001\% or less.}
\label{tab:fracvars}
\end{table}

Because of the strong correlations within the optical and with the UV data in Fig.~\ref{fig:0_LCs}, the similar fractional variability values, and the number of data points available in each band, all further analyses will be carried out using only the UV and X-ray data. 

Of the >1000 data points in each light curve only a limited number of nights contain two observations, and the rest contain only one. Both sources are known to exhibit variability timescales as short as minutes in the X-ray regime \citep{Zhang1999, Tramacere2009, Kapanadze2017_421, Kapanadze2017_501, Kapanadze2018a, Kapanadze2018b}, however no short term variability in the UV \citep{Sambruna2000, Zeng2019, Chatterjee2021}. For these reasons, we focus on only long-term variability in this paper.

\subsection{Flux distributions} \label{sec:flux_dists}

\begin{figure*}[htb]
    \centering
    \subfloat{{\includegraphics[width=0.5\linewidth]{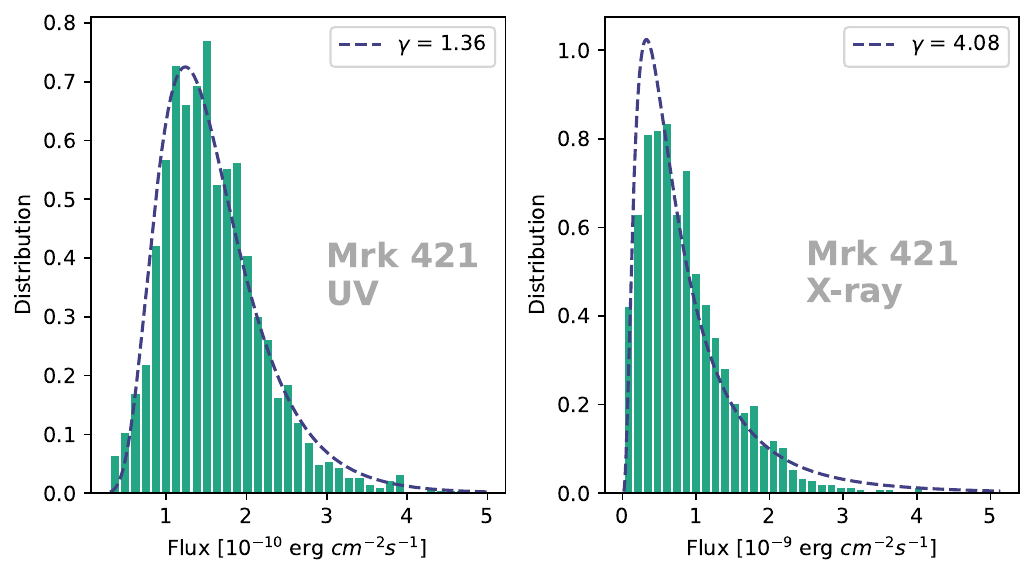}}}
    \hfill
    \subfloat{{\includegraphics[width=0.5\linewidth]{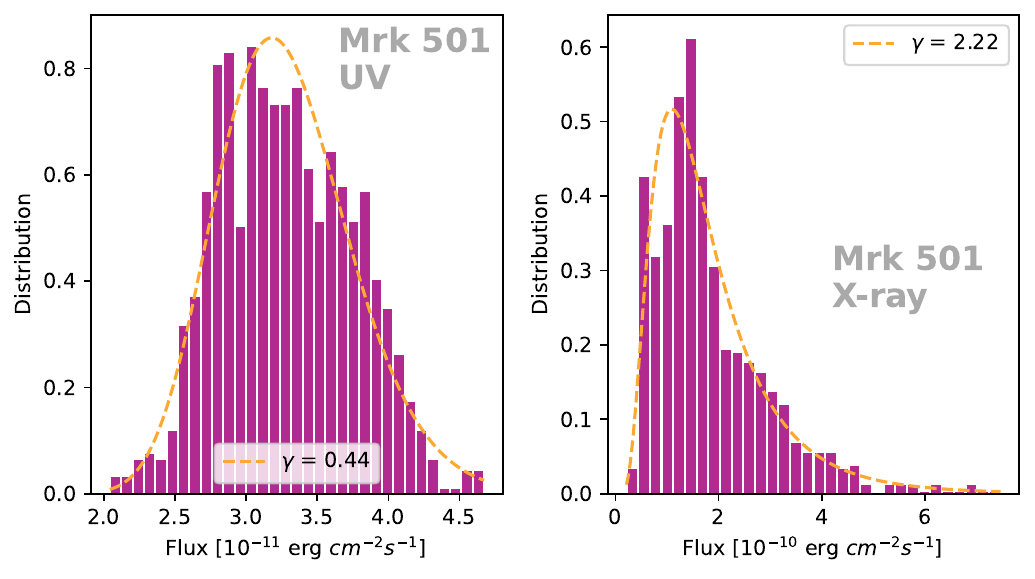}}}
    \caption{Flux distributions for, from left to right, Mrk\,421 UV, Mrk\,421 X-ray, Mrk\,501 UV, and Mrk\,501 X-ray. The skewness parameter, $\gamma$, a metric of lognormality, is shown in the legend and increases with photon energy.}
    \label{fig:1_FHs}
\end{figure*}

The flux distribution functions of the UV and X-ray flux of Mrk\,421 and Mrk\,501 are shown in the left (green) and right (purple) panels of Fig.~\ref{fig:1_FHs}, respectively. To each flux distribution we fit a lognormal distribution, which is defined as

\begin{equation}
    \mathcal{L}(x) = \frac{1}{xs\sqrt{2\pi}}
    \exp\left(-\frac{(\ln(x)-m)^2}{2s^2}\right),
\end{equation}

\noindent where $m$ (units of $\ln$(flux)) and $s$ are the mean location and the shape parameters of the distribution, respectively. The skewness parameter, $\gamma$, of a lognormal distribution is written as 

\begin{equation}
    \gamma = \left(\exp(s^2) + 2\right) \sqrt{\exp(s^2) - 1},
\end{equation}

and is presented for each fitted distribution in the figure legends. We see a positive trend of skewness with frequency band, that is, the lognormality observed is energy dependent.

The lognormal distribution of flux has been commonly observed at high energies in blazars: first in the X-ray band for BL Lacertae \citep{Giebels2009}, then in the $\gamma$-ray band for many sources \citep[e.g.,][]{HESS2010Abramowski, HESS2017Abdalla, Bhatta2020}. For Mrk\,421, lognormality has also been observed in the lower energy radio and optical bands \citep{Sinha2016}, and in the optical and UV bands \citep{Kapanadze2024}. We have found a trend of weaker lognormality toward longer wavelengths, which is in line with the results of \cite{Ciaramella2004, HESS2017Abdalla} and \cite{Kapanadze2023}. This is expected in the case of random fluctuations in the particle acceleration rate caused by small temporal fluctuations in the accretion disc \citep{Sinha2018}.

A characteristic of the lognormal process is that fluctuations in the flux are, on average, proportional to the flux itself \citep{Aitchinson1963}, that is to say the rms-flux relation is linear \citep{Uttley2001}. In blazars, these processes are often attributed to stochastic multiplicative processes \citep{Uttley2005}, and linked to the underlying accretion process of the disc on the jet \citep{Uttley2001, Giebels2009}. 
If damping is negligible, density fluctuations and instabilities in the disc, arising on local viscous timescales, can propagate inwards and couple together so as to produce a multiplicative behaviour in the accretion rate, see \cite{Rieger2019} and references therein. If this effect is efficiently transmitted into the jet, the emission could be modified accordingly \citep{HESS2017Abdalla}. Other possible multiplicative processes occurring in blazars are cascade-like events such as synchrotron radiation from the proton-induced hadronic cascades \citep{HESS2017Abdalla, Rieger2019}, or the jets-in-jets scenario \citep{Giebels2012}, in which magnetic dissipation triggers the emission of relativistic smaller jets of plasma within the bulk of the jet. It is critically important to note that this latter model reconciles the flux properties and rapid variations present in blazars at high ($\gamma$-ray) energies, but research remains to establish whether it can also explain the lognormal distributions observed at lower (optical, UV, X-ray) energies \citep{Li2017}.

Whilst all of these multiplicative mechanisms are possible, \cite{Giebels2012} and \cite{Scargle2020} showed that a lognormal distribution can in fact come also from a series of additive processes, meaning that shot noise, or the superposition of many minijets, could also explain the observed flux distributions. \cite{Bhatta2020} suggest a scenario in which both additive and multiplicative processes operate at varying degrees along the extended jet. \cite{Sinha2018} and references therein, also point out that the rapid variability seen in X-ray strongly favour mechanisms within the jet, rather than in the accretion disc.

The fact that we observe lognormal flux distributions of quite different skewnesses in different energy bands is an indication that these photons are not coming from the same emission zones, hence a multi-zone model is required to fully explain the blazar's emission processes \citep{Kapanadze2020, Kapanadze2023, Kapanadze2024}.

\subsection{Photon indices} \label{sec:indices} 

\begin{figure}[tb] 
    \centering
    \begin{subfigure}{\linewidth}
       \includegraphics[width=\linewidth]{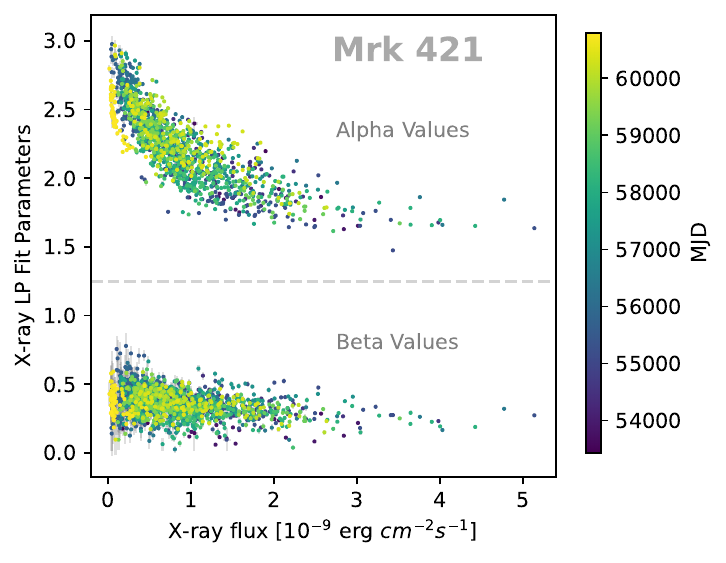}
    \end{subfigure}
    
    \begin{subfigure}{\linewidth}
       \includegraphics[width=\linewidth]{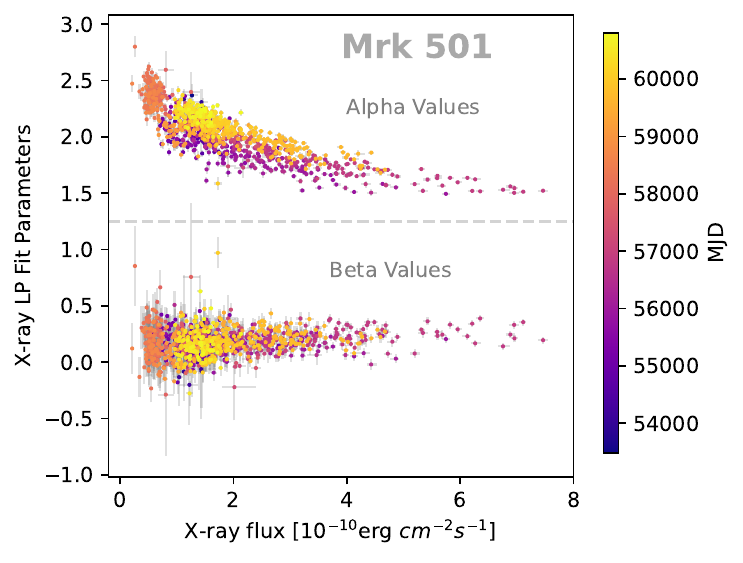}
    \end{subfigure}
    
    \caption{X-ray log parabola fit parameters, $\alpha$ and $\beta$, as a function of flux. Mrk\,421 (top) and Mrk\,501 (bottom) follow the harder-when-brighter trend, with no trend in curvature.} 
    \label{fig:2_HWBs}
\end{figure}

The values of the $\alpha$ and $\beta$ parameters used in the LP fitting of the X-ray data, as described in Eq.~\eqref{eq:logparabola}, are plotted in the top (green) and bottom (purple) panels of Fig.~\ref{fig:2_HWBs} for Mrk\,421 and Mrk\,501, respectively. Alpha values are comparable to the photon index in a power law fit, $\Gamma$ in Eq.~\eqref{eq:powerlaw}, and these terms will hereafter be used interchangeably. The X-ray photon index of Mrk\,421 varies from 1.47 to 2.98, and that of Mrk\,501 from 1.49 to 2.80. As illustrated by the colourbar, both sources fluctuate over time between hard and soft states: a hard photon spectrum has $\Gamma$ < 2, and a soft one has $\Gamma$ > 2. Moving from soft to hard spectra indicates that the synchrotron peak position shifts to higher energies and that the HBLs become `extreme' HBLs \citep{Ahnen2018, Abeysekara2020}. \cite{Malkov2001} showed that multiple shocks are required in order to produce photon indices as hard as $\sim$1.5. 

The $\beta$ parameter, that is, curvature, shows no trend with flux but does show significant variations between observations. It has an average value of $0.352\pm0.035$ for Mrk\,421, and $0.176\pm0.080$ for Mrk\,501, showing that the spectra of the former are consistently more curved than the latter. In Mrk\,501 the $\beta$ parameter falls below 0 for some data points, meaning that the spectra change shape from concave to convex, which could also be caused by a shift in synchrotron peak position. There is no correlation observed between $\alpha$ and $\beta$ in either source.

For both sources, the anticorrelation of alpha and flux is indicative of the harder-when-brighter trend seen in many blazars, including Mrk\,421 \citep{Kapanadze2020, Kapanadze2024, Alfaro2025} and Mrk\,501 \citep{Kapanadze2023, Magic2024Abe_501, Tantry2024, Alfaro2025}. This trend, that is, the hardening of the spectrum with increasing flux, can be caused by electrons that are rapidly accelerated compared to the rate at which they cool. The phases of increasing flux are when the spectra harden, and those of decreasing flux are when they soften. This flaring behaviour can be reproduced by the shock-in-jet model with escape and synchrotron losses \citep{Kirk1998}: a shock propagates through the plasma and increases the plasma's local density, which quickly involves more particles in the acceleration process \citep{Sokolov2004}.

\subsection{Periodicity search} \label{sec:Periodicity}

Studies of periodicities, or more likely to be present, QPOs, can be used to infer the physical dimensions, masses, and geometrical structures of emission zones \citep{Ghisellini2010}. In order to thoroughly investigate any potential QPOs present in the UV and X-ray light curves of Mrk\,421 and Mrk\,501, we employed three well-known techniques: the structure function (SF), the Lomb-Scargle periodogram (LSP), and the discrete auto-correlation function (DACF).

\begin{figure*}[htb]
    \centering
    \subfloat{{\includegraphics[width=0.25\linewidth]{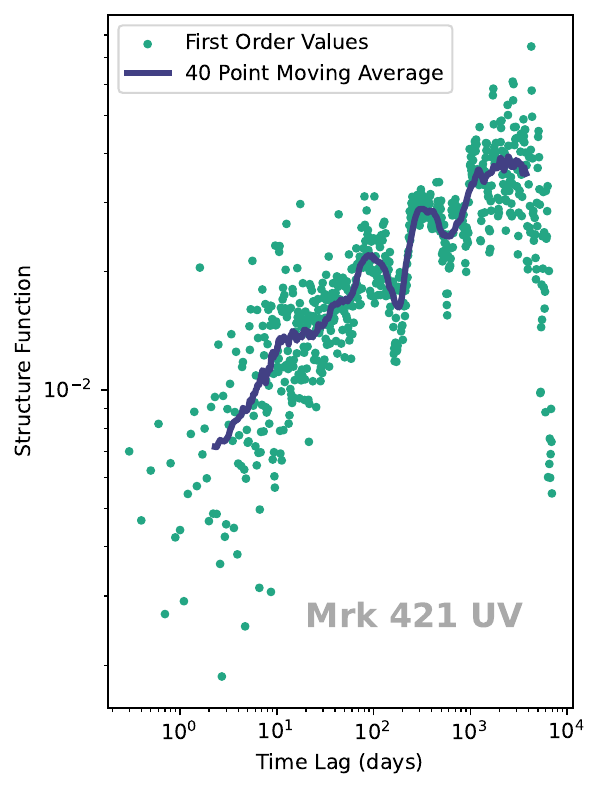}}}
    \hfill
    \subfloat{{\includegraphics[width=0.25\linewidth]{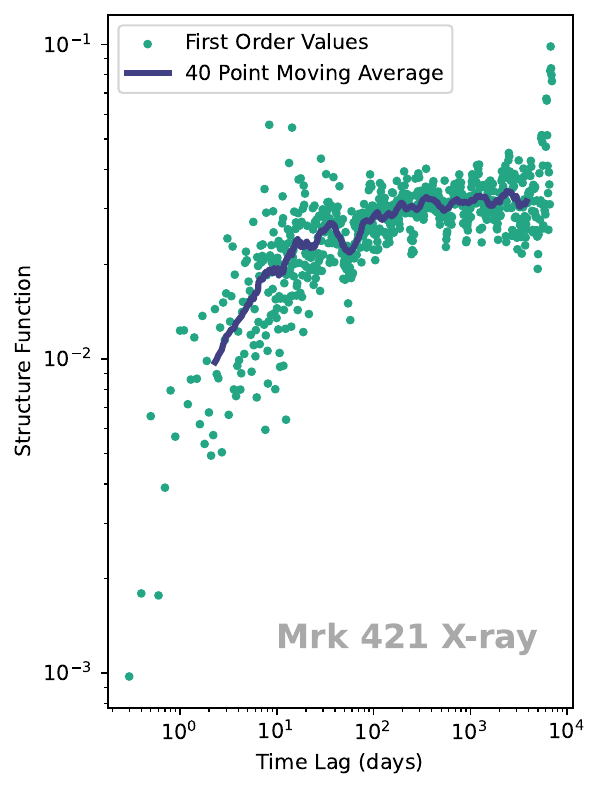}}}
    \hfill
    \subfloat{{\includegraphics[width=0.25\linewidth]{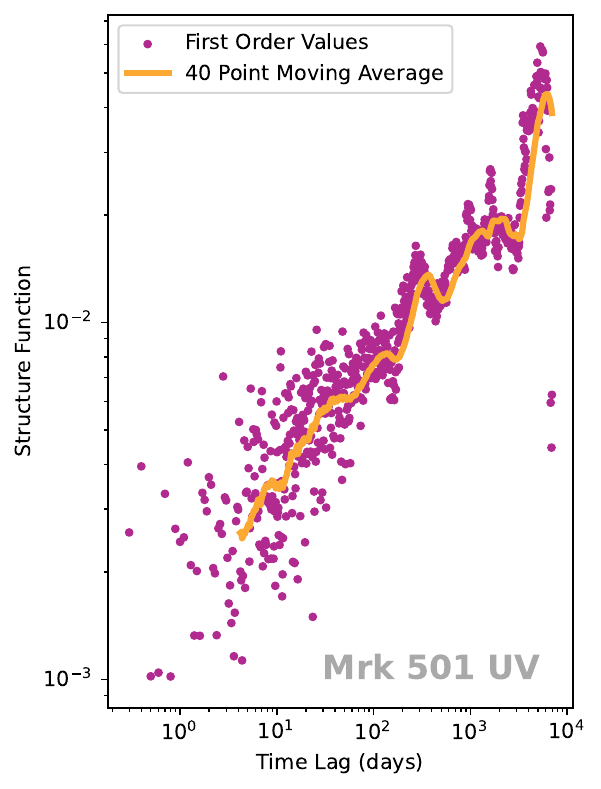}}}
    \hfill
    \subfloat{{\includegraphics[width=0.25\linewidth]{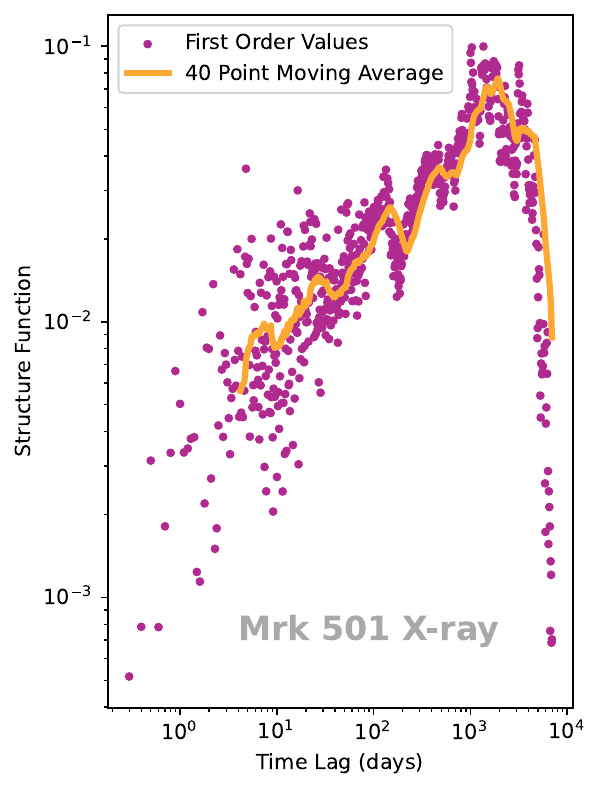}}}
    \caption{Structure functions for Mrk\,421 UV (far left), Mrk\,421 X-ray (centre left), Mrk\,501 UV (centre right), and Mrk\,501 X-ray (far right). After 2000 days these are badly affected by systematics, so peaks right of this value should not be trusted. Only Mrk\,421 X-ray appears to reach saturation.}
    \label{fig:4_SFs}
\end{figure*}

The first-order structure function \citep{Rutman1978, Simonetti1985} is defined as:

\begin{equation}
    \text{SF}(\tau) = \frac{1}{N}\sum_{i=1}^N 
                        \left[X(t_i) - X(t_i + \tau)\right]^2,
\end{equation}

\noindent where $X(t_i)$ is the flux measured at time, $t_i$, the total number of pairs of measurements is $N$, and $X(t_i)$ and $X(t_i + \tau)$ are pairs of flux measurements separated by time lag, $\tau$. The SF has the advantage of being less affected by windowing effects, often seen in blazar light curves due to the large gaps between observation periods, compared to direct Fourier methods \citep{HESS2010Abramowski}. Interpretation of the SF is an intricate deed, and commonly the break seen at longer time lag values, for example around 2000 days for Mrk\,501 in Fig. \ref{fig:4_SFs}, is wrongly accredited to being the characteristic timescale of the source (the QPO). \cite{Emmanoulopoulos2010} show that it in fact represents the characteristic timescale of the light curve: it is the maximum probeable timescale given the length of the observations. SFs can therefore be used to provide upper limits to timescale investigations such as our own. 
Since each of the light curves are slightly longer than 7000 days, timescales longer than 1400 days are  probed with less than five independent samples and hence cannot be explored with reliable statistical significance. We thus set an upper limit for probing timescales with the LSP and DACF at $\sim$1400 days.
The SFs for Mrk\,421 and Mrk 501 both using UV and X-ray data are displayed in Fig.~\ref{fig:4_SFs}. All SF decay in a power law fashion towards short timescales. The break usually seen at lower time lag values, which represents the timescale where average variability amplitudes are comparable to the average measurement errors, is not observed in our work due to the cadence of the light curves (see Sec.~\ref{sec:frac_var}). Both SFs of Mrk 501 rise in a power law fashion up to breaks beyond the longest time scales that can be probed (about 1400 days). The X-ray SF of Mrk\,421  breaks on timescales of 30-100 days with a significant change in the power law slope displayed over more than one order of magnitude towards shorter and longer timescales. A similar break is indicated in the SF of UV variations of Mrk 421, but is only marginally significant. Beyond 3000 days both SFs of Mrk 421 display further deviations form a power-law SF, but at these timescales, they no longer provide statistically meaningful information. 

Various peaks and troughs are seen in the SFs in Fig.~\ref{fig:4_SFs}, perhaps most clearly in Mrk\,421 UV, but in all cases these are found at multiples of half a year, and should therefore not be interpreted as physical timescales associated with the blazars.

The LSP \citep{Lomb1976, Scargle1982} is defined as: 

\begin{equation}
    \begin{aligned}
    P(f) = \frac{A^2}{2}\left( \sum_n g_n\cos(2\pi f[t_n-\tau]\right)^2 
    \\
    \text{}+ \frac{B^2}{2}\left( \sum_n g_n\sin(2\pi f[t_n-\tau]\right)^2,
    \end{aligned}
\end{equation}

\noindent where $A,B$, and $\tau$ are arbitrary functions of the frequency $f$ and observing times \{$t_i$\} \citep{VanderPlas2018}. Incorporating Fourier analysis, the least-squared method, and Bayesian probability theory, the LSP's strength lies in its ability to examine QPOs in unevenly sampled time series. The standard LSP does not account for measurement errors, and assumes that the fitted sine function has the same mean as the data, hence we use the generalised LSP \citep{Zechmeister2009} to avoid these shortcomings. In order to assess the validity of peaks in the LSP, we employ the false alarm probability (FAP) method \citep{Baluev2008} and the Monte Carlo method. For the latter, we generate 10,000 synthetic light curves with the same power spectral density, probability density function, and sampling as the observed light curves, following the \cite{Emmanoulopoulos2013} algorithm implemented in the 
gammapy\textunderscore SyLC package\footnote{\url{https://github.com/cgalelli/gammapy\textunderscore SyLC}}, and necessitate a significance of $\geq3\sigma$ \citep{Vaughan2005}. Matching the cadence of the observations in the synthetic light curves intrinsically accounts for any aliases that would occur due to windowing.

\begin{figure*}[ht]
  \begin{subfigure}[b]{0.5\linewidth}
    \centering
    \includegraphics[width=\linewidth]{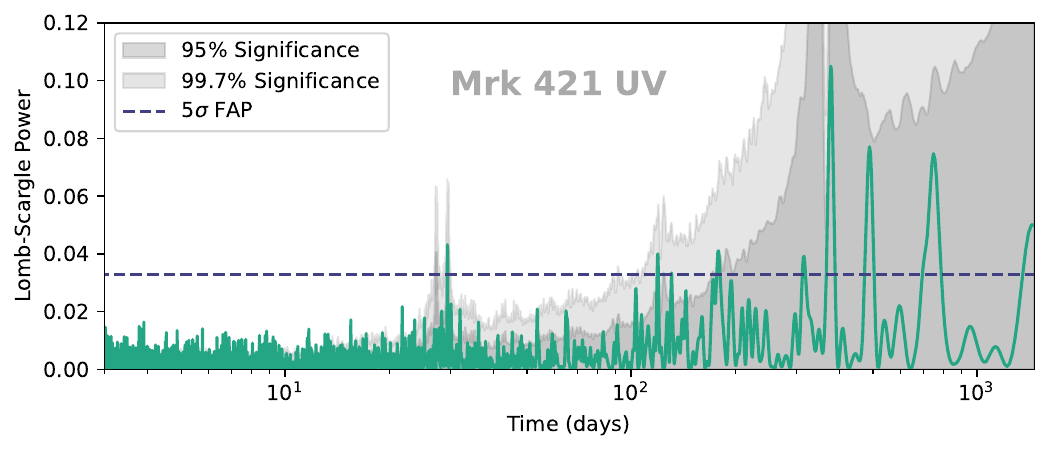}
  \end{subfigure}
  \begin{subfigure}[b]{0.5\linewidth}
    \centering
    \includegraphics[width=\linewidth]{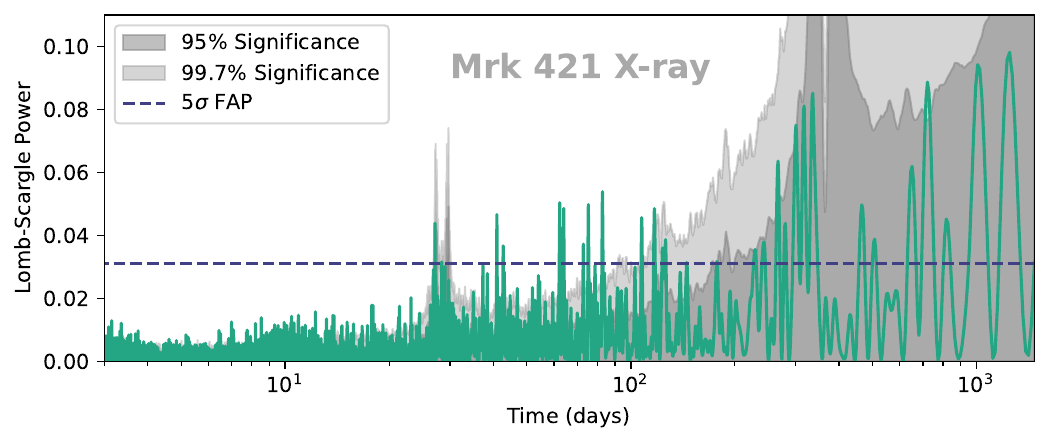}
  \end{subfigure}
  \begin{subfigure}[b]{0.5\linewidth}
    \centering
    \includegraphics[width=\linewidth]{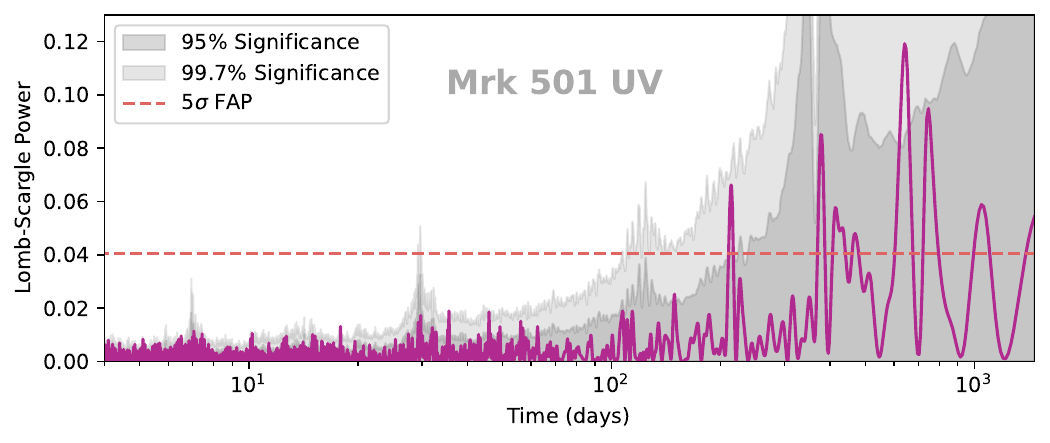}
  \end{subfigure}
  \begin{subfigure}[b]{0.5\linewidth}
    \centering
    \includegraphics[width=\linewidth]{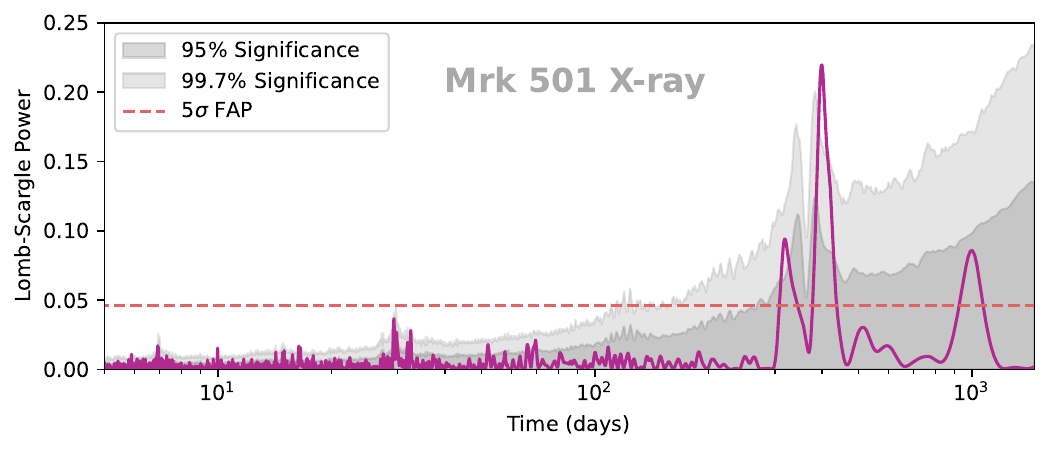}
  \end{subfigure}
  \caption{Lomb-Scargle periodograms for Mrk\,421 UV (top left), Mrk\,421 X-ray (top right), Mrk\,501 UV (bottom left), and Mrk\,501 X-ray (bottom right). Only the latter contains a peak above the $3\sigma$ significance level, which falls at around 400 days.}
  \label{fig:4_LSPs}
\end{figure*}

The LSPs for Mrk\,421 UV and X-ray, and Mrk\,501 UV and X-ray, are displayed in Fig.~\ref{fig:4_LSPs}. There are no potential QPOs detected in the UV light curves of Mrk\,421 or Mrk\,501. The X-ray LSP of Mrk\,421 shows 11 peaks between 37 and 125 days, but after careful inspection of the peaks in the same time frame for the synthetic light curves we conclude that these QPOs should not be interpreted as genuine. This is because many peaks appear in the same range of time lag values in almost all of the synthetic LSPs, but are not reflected in the grey significance bands due to slight shifts in the exact time at which each peak appears. 
There is a potential QPO detected above the required 3$\sigma$ significance level and the 5$\sigma$ FAP in the X-ray LSP of Mrk\,501 at $400_{-17}^{+24}$ days. The uncertainties here have been estimated by fitting a skewed Gaussian and calculating its two half width at half maximum values. These results will be discussed further in this section, after the DACF results.

\begin{figure*}[ht] 
  \begin{subfigure}[b]{0.5\linewidth}
    \centering
    \includegraphics[width=\linewidth]{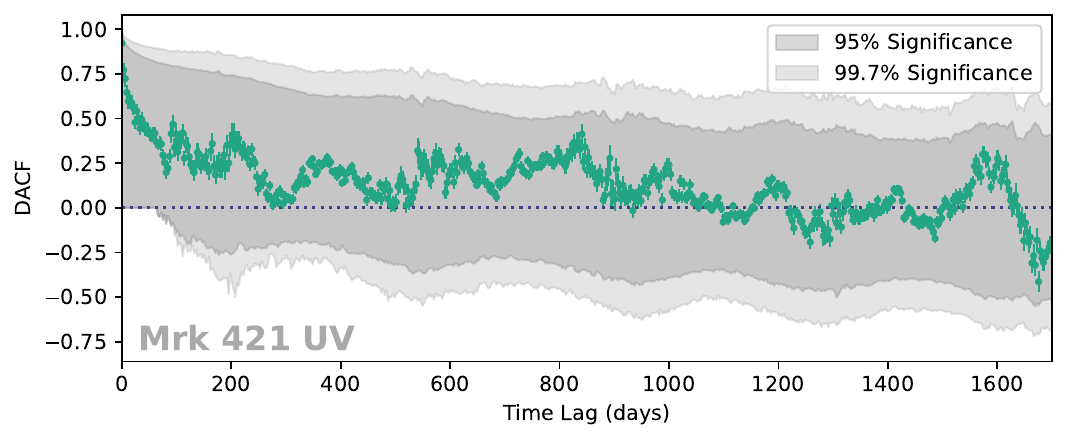} 
  \end{subfigure}
  \begin{subfigure}[b]{0.5\linewidth}
    \centering
    \includegraphics[width=\linewidth]{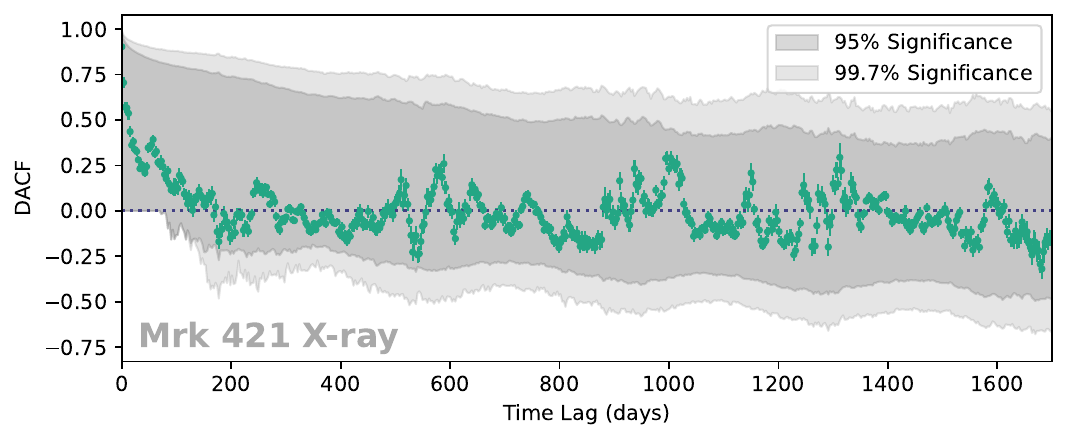} 
  \end{subfigure} 
  \begin{subfigure}[b]{0.5\linewidth}
    \centering
    \includegraphics[width=\linewidth]{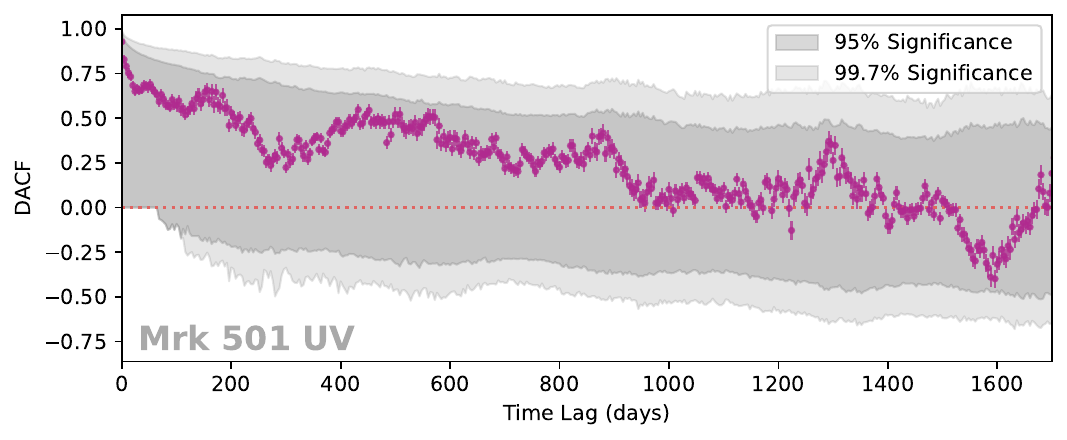} 
  \end{subfigure}
  \begin{subfigure}[b]{0.5\linewidth}
    \centering
    \includegraphics[width=\linewidth]{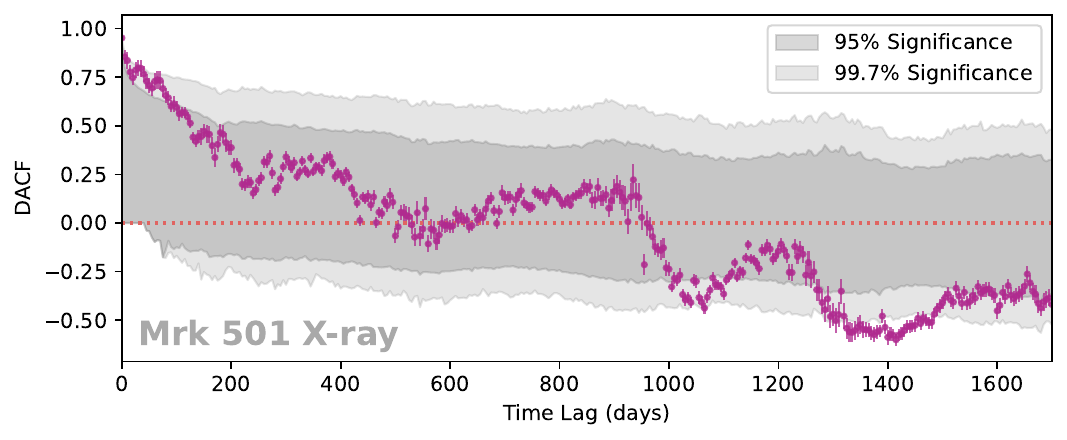} 
  \end{subfigure} 
  \caption{Discrete auto-correlation functions for Mrk\,421 UV (top left), Mrk\,421 X-ray (top right), Mrk\,501 UV (bottom left), and Mrk\,501 X-ray (bottom right). Again only the latter contains any points above the $3\sigma$ significance level, at around 1400 days.} 
  \label{fig:4_DACFs} 
\end{figure*}

The third approach we use to investigate possible QPOs was the DACF, which is a special case of the discrete correlation function \citep[DCF,][]{EdelsonKrolik1988} where the same time series is correlated with itself. The DCF is defined as

\begin{equation}
    \text{DCF}(\tau) = \frac{1}{N} \sum_{i,j} 
                        \frac{(a_i - \bar{a}) (b_i - \bar{b})}
                             {\sigma_a \sigma_b},
    \label{eq:DCF}
\end{equation}

\noindent where $a$ and $b$ are the values (in our case, fluxes) of the time series, with mean values $\bar{a}$ and $\bar{b}$, and standard deviations $\sigma_a$ and $\sigma_b$, respectively. $N$ is the number of pairs $i,j$ in the time interval $\tau$. In the DACF, $a=b$, and we are able to investigate timescales intrinsic to the light curve. We generated synthetic light curves following the same structure as outlined for the LSPs to create significance contours for the DACFs. Due to the colour noise nature of blazars \citep{Press1978, Lawrence1987} we expect the DACF value to slowly decrease as time lag increases, representing the inherent autocorrelation present in these sources. Points of interest are therefore: the time lag value at which this decreasing trend breaks, how clearly it breaks, and any distinct peaks (or troughs) exceeding the 3$\sigma$ significance bands.

The DACFs for Mrk\,421 UV and X-ray, and Mrk\,501 UV and X-ray, are displayed in Fig.~\ref{fig:4_DACFs}. 
In Mrk\,421 UV and X-ray, breaks appear at $\sim$84 days and $\sim$42 days, respectively, where the latter of these results is a harmonic of the former, indicating a possibly genuine QPO. After roughly 200 days both Mrk\,421 DACFs begin to fluctuate around zero, indicating no autocorrelation after this time lag. The DACFs of Mrk\,501 UV and X-ray show a variety of features between 25 and 280 days, included in both is a peak at roughly 200 days, but this is not $>3\sigma$ significant. No significant features appears to be present at $\sim$400 days in the Mrk\,501 UV or X-ray DACFs, where the potential QPO lay in the corresponding LSP, but it is worth mentioning that the 200 day peak corresponds to a harmonic of the 400 day LSP peak.
Also interesting to note is the $>3\sigma$ trough in the Mrk\,501 DACF at $\sim$1400 days: for a time series with period $T$, a negative autocorrelation value occurs when the time lag examined is out of phase with the dominant peak in the light curve, that is $(n+1/2)T$. If $T=400$ days from the LSP, then the trough falls exactly at $n=3$. At roughly $n=2$, or a time lag of 1000 days, is another trough which falls just below $3\sigma$ significance.

At $n=1$ (600 days) the strongest aliasing feature would be expected, but is not observed. 
In the Mrk\,501 UV LSP in Fig.~\ref{fig:4_LSPs} we see a peak at $\sim$400 days, and peaks corresponding to harmonics at $\sim$200 and $\sim$600 days that are $>2\sigma$ significant.

There is lack of consistency between the LSPs and DACFs, as well as within the DACFs themselves, e.g. why do troughs appear at $n=2$ and $n=3$, but not at $n=1$, and why do no significant peaks ($nT$) appear at any value of $n$? These questions do not strengthen the indication of any QPO results. Mainly, they highlight the necessity for more densely sampled light curves in blazar analyses.

To the best of our knowledge, currently no QPOs have been reported for Mrk\,421 in the UV or optical bands, despite efforts by e.g. \cite{Carnerero2017, Sandrinelli2017, Tarnopolski2020, Kapanadze2020, Otero-Santos2023, Kapanadze2024}, and \cite{Abe2025}, all of whom also found no QPOs in any other band. The latter statement is contradictory to the work of e.g. \cite{Li2016, Bhatta2020}, and \cite{Ren2023}, who found QPOs ranging from 278 - 310 days at radio, X-ray, and/or gamma-ray energies.

Over the years, QPOs in Mrk\,501 have been reported at varying timescales: $\sim$23 days, or harmonics thereof, by \cite{Hayashida1998, Protheroe1998, Kranich1999, Nishikawa1999}, and \cite{Roedig2009} in the X-ray and/or gamma-rays; and 330 and 333 days in the gamma and the optical by \cite{Bhatta2019} and \cite{Bhatta2021}, respectively. Possible mechanisms responsible for the $\sim$23 day QPO has been investigated thoroughly \citep{Rieger2000, Rieger2005, Yu-Hai2005, Fan2008}. Both \cite{Devanand2022} and \cite{Kapanadze2023} found no QPOs at UV, X-ray, and/or gamma-ray energies. We believe this is the first time a possible QPO of $\sim$400 days is reported for Mrk\,501. Because 400 days is a year-like timescale, we have taken care to check that this peak is not just a consequence of windowing, nor does it appear in the Mrk\,501 UV LSP if only the data with matching MJD values (within 0.1 days) to the Mrk\,501 X-ray light curve are used.

In its host frame, the reported QPO in Mrk\,501 corresponds to $390^{+23}_{-16}$ days, using $P'=P_{obs}/(1+z)$. We offer three possible explanations for this observation. The first is a supermassive binary black hole (SMBBH) system in which the less massive black hole emits nonthermal radiation via its relativistic jet \citep{Valtonen2011}. The jet exhibits orbital motion around the system's centre of mass, which means the Doppler factor for the emission region is a periodic function of time \citep{Rieger2000}. This model can also produce complex QPOs due to mergers and other such interactions \citep{Valtonen2008}. Whilst SMBBHs are possible, the number of year-like physical periods ($P'$) detected in blazars, for example in this work for Mrk\,501, are unlikely to all be caused by such systems due to gravitational constraints \citep{Rieger2019}. 

The second explanation stems from the helical motion of a component inside a jet, which itself is rotating, where differential Doppler boosting explains periodic lighthouse effects \citep{Camenzind1992}. This model is closely linked to the model of \cite{Schramm1993} and has a dynamical timescale of days to years. However, in such models long timescale QPOs do not complete many oscillations, hence their signal is drowned out. We therefore cannot credit this mechanism with the production of the 390 day (host frame) QPO we observe.

The third scenario is concerned with instabilities intrinsic to the accretion disc \cite{McHardy2006}, such as bright hot-spots orbiting on the disc \citep{Bhatta2018}, Lense-Thirring precession \citep{Stella1998}, and jet precession due to a warped accretion discs \citep{Liska2018}. Of these accretion disc scenarios, the first two can be on the same timescales as the QPOs found in this work, but the latter cannot, expecting $\sim10^3$ years \citep{Liska2018}. \cite{Chatterjee2018} point out that if it is the accretion disc which regulates the jet emission, or creates the shocks which move through the jet, then QPOs found in any energy band should be the same for the same source. This means that all reported QPOs should be close in value. If we believe that our QPO is a result of the imprinting of the disc onto the jet, this is also supported by the lognormality observed in the flux distributions in Sec.~\ref{sec:flux_dists}, which can be explained in the same way \citep{Uttley2001}. Similarly, \cite{ArbetEngels2021_501} found a characteristic time between TeV flares in Mrk\,501 of 5-25 days, which is comparable to the timescale associated with flares triggered by Lense-Thirring precession of the accretion disc. Together these results provide evidence for a disc-jet connection in Mrk\,501, despite the lack of directly observed disc features.

The lack of detections of QPOs in Mrk\,421 UV and X-ray and Mrk\,501 UV are still valuable in that they provide lower limits to any QPOs that may exist in these objects at these energies. That is, both sources may have genuine QPOs that are simply longer than timescales we are able to reasonably probe: $>$1400 days. Of course, we have also limited our search to QPOs longer than 3-5 days due to sampling constraints. Thus, an upper limit to QPO period is also imposed. For a full discussion of factors limiting QPO detections see \cite{VaughanUttley2005}.

\subsection{Cross correlations} \label{sec:CrossCorrs}

As mentioned in Sec.~\ref{sec:LCs}, simultaneous correlations between optical and UV bands, like those seen in Fig.~\ref{fig:3_Corrs}, are expected due to the broad-band emissivity of photons in that energy range. This has frequently been confirmed to be true, for example by \cite{Kapanadze2016, Kapanadze2017_421, Kapanadze2017_501, Kapanadze2018a, Kapanadze2018b,Kapanadze2020, Kapanadze2023, Kapanadze2024, Abe2025}. All of these also found no or only weak long-term simultaneous correlations between optical/UV and X-ray bands, as we have in Sec.~\ref{sec:LCs}.

\begin{figure}[tb] 
    \centering
    \begin{subfigure}{\linewidth}
       \includegraphics[width=\linewidth]{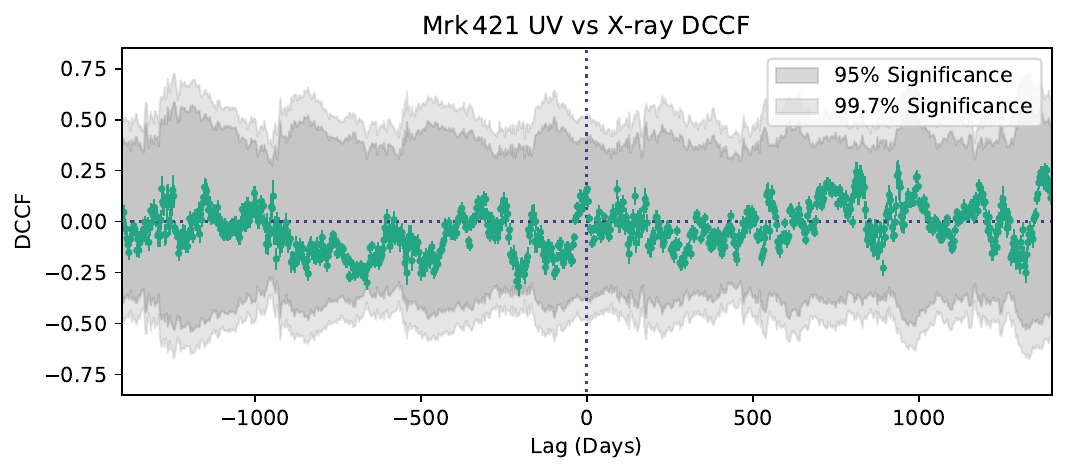}
    \end{subfigure}
    
    \begin{subfigure}{\linewidth}
       \includegraphics[width=\linewidth]{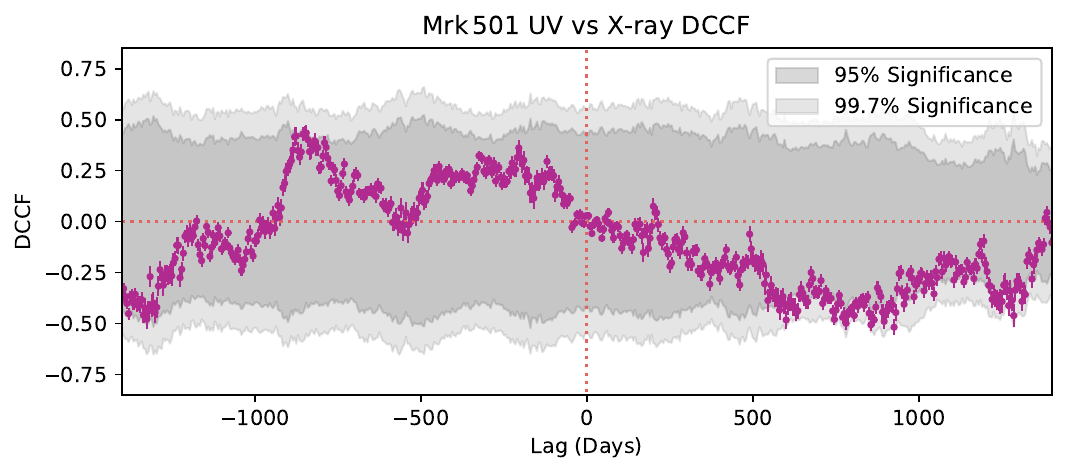}
    \end{subfigure}
    
    \caption{The UV vs X-ray discrete cross-correlation functions for Mrk\,421 (top) and Mrk\,501 (bottom).} 
    \label{fig:5_DCCFs}
\end{figure}

Moreover, we have used the discrete cross-correlation function (DCCF), defined in Eq.~\eqref{eq:DCF} with $a\not=b$, to investigate if the optical/UV bands lead or lag behind the X-ray band. The length of our light curves allow us to probe longer than ever lag ranges. We limit this study to lags between -1400 and 1400 for the same reasons as given in Sec.~\ref{sec:Periodicity}. The UV vs X-ray DCCFs for Mrk\,421 and Mrk\,501 are shown in the top and bottom panel of Fig.~\ref{fig:5_DCCFs}, respectively, with the grey significance bands generated in exactly the same way as with the DACFs in Sec.~\ref{sec:Periodicity}. Neither source shows any significant correlation at any given lead or lag, by either band. We cannot however exclude cross correlations on timescales of a day or less, for the same sampling reasons explained in Secs.~\ref{sec:frac_var} and \ref{sec:Periodicity}.

When focussing on shorter defined periods of specific flares \cite{Kapanadze2016, Kapanadze2017_421, Kapanadze2018b, Kapanadze2020, Kapanadze2023}, and \cite{Kapanadze2024} found weak but statistically significant simultaneous positive correlations between the X-ray and optical/UV bands. In fact, in \cite{Kapanadze2018b} and \cite{Kapanadze2020} these correlations are also negative in some periods. Using the entire period of our light curves we confirm that, whilst during certain periods a quasi-simultaneous X-ray - optical/UV correlation may be present, overall it is not. This gives further evidence for the multi-zone model, and even suggests that the number of zones is not constant but changing, hence why sometimes correlations do appear, but not consistently.

Given that both sources are classified as HBLs, the X-ray emission falls within the lower energy peak in the SED \citep{Abdo2010}, which implies an optical/UV - X-ray correlation under the one-zone model. Recent work on X-ray polarisation data by \citep{Liodakis2022} suggests a shock front in the jet as the source of particle acceleration, which also insinuates optical/UV - X-ray correlation. Conversely, the lack of correlation observed in this work, and in work by \cite{Kapanadze2020, Kapanadze2023, Kapanadze2024, Magic2024Abe_501, Abe2025}, suggests a multi-zone model. Observations and modelling of an intriguing case of a blazar showing evidence for the multi-zone model can be found in \cite{HESS_Aharonian2023}.

\section{Conclusions} \label{sec:concs}

The HBLs Mrk\,421 and Mrk\,501 have been observed for 20 years, from 2005 to 2025, with the Swift-UVOT and Swift-XRT telescopes, providing us with rich MWL datasets and valuable insights into the nature of blazars. We investigated long-term trends and summarise our key findings as follows:

\begin{enumerate}
    \item The Swift X-ray spectra studied in this work are statistically better fit with a log parabola model than a single power law.
    \item Subpopulations of the X-ray data, defined by hard and soft spectral states, show an increase in correlation with the UV data, in contrast to that of the whole population.
    \item Fractional variability is dependent on energy for both sources. This could be due to the cooling times of the electron populations which produce the respective photons \citep{Tantry2024, Abe2025}.
    \item The fluxes follow a lognormal distribution, but much more pronouncedly in the X-rays than in the optical/UV, for both blazars. Lognormal distributions may be created by additive or multiplicative processes (or a combination of the two) \citep{Uttley2001, Sinha2018, Scargle2020}. The energy dependence of the distributions can be explained by fluctuations in the particle acceleration rate, which begin as fluctuations in the accretion disc \citep{Sinha2018}. The differing distribution shapes are supportive of a multi-zone model.
    \item X-ray photon indices vary with flux following a general harder-when-brighter trend, signifying that the rate at which electrons are accelerated is much faster than the rate at which they cool. This flaring activity indicates a rapid variability in the number of electrons producing X-ray photons. 
    \item The X-ray photon indices also shift from hard to soft and back again many times over the 20 years of observations, reflecting the phases of flux increases (spectral hardening), and flux decreases (spectral softening). This means that shifts in the position of the synchrotron SED peak have occurred.
    \item An in-depth time series analysis has revealed a potential QPO of $\sim$390 days (host frame) for Mrk\,501 in the X-ray. QPOs on this timescale can be explained by accretion disc instabilities such as hot spots and Lense-Thirring precession \citep{Stella1998, Bhatta2018}. The lack of detections of QPOs in Mrk\,421 UV and X-ray, and Mrk\,501 UV, imply QPO limits of <3 days or >1400 days and <4 days or >1400 days, respectively
    \item Strong optical - UV correlations were observed at zero time lag, but none were for optical/UV - X-ray correlations in any lag between -1400 and 1400 days. This contradicts the implications of a one-zone model, in which all wavelengths within the lower SED peak should be well correlated. Once again, giving evidence for a multi-zone emission model.
\end{enumerate}

Overall, this work gives a comprehensive analysis of the temporal characteristics of Mrk\,421 and Mrk\,501 in the UV and X-ray bands as recorded with Swift. These results strongly favour a multi-zone model to explain the emissions of Mrk\,421 and Mrk\,501 over the last 20 years.

\begin{acknowledgements}  
We thank the anonymous referee for their constructive feedback, which helped to improve the manuscript. We also wish to thank Frank Rieger, Sarah Wagner, and Daniela Dorner for interesting and very helpful discussions at the DFG RU 5195 Annual Assembly 2025 and elsewhere. 
We gratefully acknowledge funding by the Deutsche Forschungsgemeinschaft (DFG, German Research Foundation), within the research unit 5195 ``Relativistic Jets in Active Galaxies'' -- project number 443220636.
The project on which this report is based was partially funded by the Bundesministerium für Bildung und Forschung (BMBF, Ministry of Education and Research). Responsibility for the content of this publication lies with the author.
The project is co-financed by the Polish National Agency for Academic Exchange.
The authors gratefully acknowledge the Polish high-performance computing infrastructure PLGrid (HPC Center: ACK Cyfronet AGH) for providing computer facilities and support within computational grant no. \text{PLG/2024/017925}. 
This work is funded in parts by the Deutsche Forschungsgemeinschaft (DFG, German Research Foundation) -- project number 460248186 (PUNCH4NFDI).
\end{acknowledgements}

\bibliography{bib}

\appendix

\onecolumn
\section{Correlation plots}\label{app:CorrsPlots}

\begin{figure*}[ht!]
    \centering
     \resizebox{17.6cm}{22cm}
    {\includegraphics {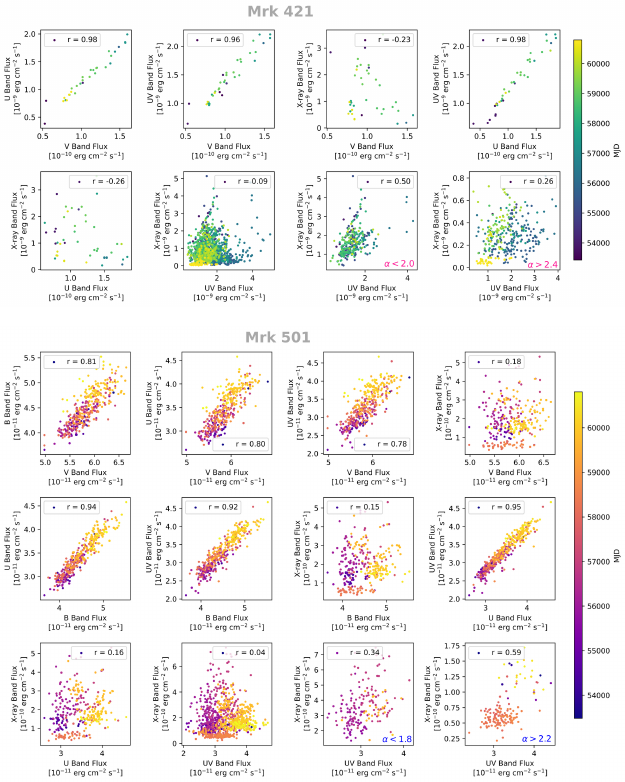}}
     \caption{Correlations between energy bands for Mrk\,421 (top) and Mrk\,501 (bottom), where $r$ is the Pearson correlation coefficient. There are strong correlations within the optical and UV bands. If considering the entire period there are no correlations between the optical/UV and X-ray bands, but some do appear when the X-rays are divided into hard and soft states.}
      \label{fig:3_Corrs}
\end{figure*}

\end{document}